\documentclass[10pt,letterpaper]{article}
\usepackage[top=0.85in,left=2.75in,footskip=0.75in]{geometry}

\usepackage{amsmath,amssymb}

\usepackage{changepage}

\usepackage[utf8x]{inputenc}

\usepackage{textcomp,marvosym}

\usepackage{cite}

\usepackage{nameref,hyperref}

\usepackage[right]{lineno}

\usepackage{microtype}
\DisableLigatures[f]{encoding = *, family = * }

\usepackage[table]{xcolor}

\usepackage{array}

\newcolumntype{+}{!{\vrule width 2pt}}

\newlength\savedwidth

\raggedright
\usepackage[aboveskip=1pt,labelfont=bf,labelsep=period,justification=raggedright,singlelinecheck=off]{caption}

\makeatletter
\renewcommand{\@biblabel}[1]{\quad#1.}
\makeatother

\date{}

\usepackage{lastpage,fancyhdr,graphicx}
\usepackage{epstopdf}
\fancyheadoffset[L]{2.25in}
\fancyfootoffset[L]{2.25in}
\newcommand{\x}{x_{\mathrm{Na}}}

\begin{document}
\vspace*{0.2in}

\begin{flushleft}
{\Large
\textbf\newline{Ultrafast population coding and axo-somatic compartmentalization} % Please use "title case" (capitalize all terms in the title except conjunctions, prepositions, and articles).
}
\newline
\\
Chenfei Zhang\textsuperscript{1,2,3\Yinyang},
David Hofmann\textsuperscript{4,5\Yinyang},
Andreas Neef\textsuperscript{1,2*},
Fred Wolf\textsuperscript{1,2*}
\\
\bigskip
\textbf{1} Department for Nonlinear Dynamics, Max Planck for Dynamics and Self-Organization, Göttingen, Germany
\\
\textbf{2} Bernstein Center for Computational Neuroscience, Göttingen, Germany,
\\
\textbf{3} Graduate School for Neurosciences, Biophysics, and Molecular Biosciences, University of Göttingen, Göttingen, Germany
\\
\textbf{4} Department for Physics, Emory University, Atlanta, GA, USA
\\
\textbf{5} Initiative in Theory and Modeling of Living Systems, Emory University, Atlanta, GA, USA
\\
\bigskip

% Insert additional author notes using the symbols described below. Insert symbol callouts after author names as necessary.
% 
% Remove or comment out the author notes below if they aren't used.
%
% Primary Equal Contribution Note
\Yinyang These authors contributed equally to this work.

% \textcurrency b Insert second current address 
% \textcurrency c Insert third current address

% Use the asterisk to denote corresponding authorship and provide email address in note below.
* fred@nld.ds.mpg.de
* aneef@gwdg.de

\end{flushleft}
\section*{Abstract}
Cortical neurons in the fluctuation driven regime can realize ultrafast population encoding. The underlying biophysical mechanisms, however, are not well understood. Reducing the sharpness of the action potential onset can impair ultrafast population encoding, but it is not clear whether a sharp action potential onset is sufficient for ultrafast population encoding. One hypothesis proposes that the sharp action potential onset is caused by the electrotonic separation of the site of action potential initiation from the soma, and that this spatial separation also results in ultrafast population encoding. Here we examined this hypothesis by studying the linear response properties of model neurons with a defined initiation site. We find that placing the initiation site at different axonal positions has only a weak impact on the linear response function of the model. It fails to generate the ultrafast response and high bandwidth that is observed in cortical neurons. Furthermore, the high frequency regime of the linear response function of this model is insensitive to correlation times of the input current contradicting empirical evidence. When we increase the voltage sensitivity of sodium channels at the initiation site, the two empirically observed phenomena can be recovered. We provide an explanation for the dissociation of sharp action potential onset and ultrafast response. By investigating varying soma sizes, we furthermore highlight the effect of neuron morphology on the linear response. Our results show that a sharp onset of action potentials is not sufficient for the ultrafast response. In the light of recent reports of activity-dependent repositioning of the axon initial segment, our study predicts that a more distal initiation site can lead to an increased sharpness of the somatic waveform but it does not affect the linear response of a population of neurons.

% Please keep the Author Summary between 150 and 200 words
% Use first person. PLOS ONE authors please skip this step. 
% Author Summary not valid for PLOS ONE submissions.   
%\section*{Author Summary}

%\linenumbers

% Use "Eq" instead of "Equation" for equation citations.
\section*{Introduction}
% general intro: role of AP
Neurons communicate via brief, all-or-none action potentials (APs) that are initiated close to the cell body, at the axon initial segment (AIS), and propagate along the axon. Depending on the degree of axonal branching, a single neuron can contact tens of thousands downstream neurons, and at every contact point the action potential is converted to a transient conductance change in the recipient cells. Central neurons receive thousands of such inputs every second but only a tiny fraction of this incoming information is encoded in the timing of their action potentials outputs. The precise timing of an action potential determines which aspects about the input it can encode. This action potential timing is ultimately decided by the mechanisms affecting the initiation of the action potential.

The dynamics of AP initiation is controlled by the voltage dependence of all conductances in the proximal axon that are active in the vicinity of the threshold voltage of action potential initiation. Indeed a cortical neuron's membrane potential is typically close to threshold and so operates in the fluctuation driven regime~\cite{Softky1993}. In particular the voltage dependence of the sodium channels has been identified as the key determinant of the AP onset dynamics~\cite{fourcaud-trocme_how_2003}, reflected in two different measurables: i)~The slope of the phase space plot ($\dot{V}(t)$ vs $V(t)$) at the AP onset, often termed onset rapidness. ii)~The precision of action potential timing or, in the frequency domain, the bandwidth of input frequencies that exert an influence over this timing.\\ %In experiments with pyramidal cells in brain slices and in-vivo, both measures have been found to be higher than expected from biophysically motivated conductance-based neuron models. The onset rapidness of the somatic action potential waveform reaches $30 \mathrm{ms}^{-1}$ and the bandwidth is typically $200-400 \mathrm{Hz}$ or even $1 \mathrm{kHz}$ in the case of human cells~\cite{higgs_conditional_2009,kondgen_dynamical_2008,naundorf_unique_2006,tchumatchenko_ultrafast_2011,testa-silva_high_2014}.
% why is rapidness/sharpness important
While the exact shape of the action potential waveform at its onset could be considered irrelevant for neuronal information processing, the spectral bandwidth of the encoder, that is a population of neurons, turns out to be of great importance to the function of neuronal networks. It defines an upper limit to the information transfer between neurons\cite{Lindner2001,Monteforte2010}. Incoming action potentials could potentially convey sub-millisecond timing since postsynaptic potentials rise within less than a millisecond, thus the transmission at a chemical synapse can be so reliable that it adds no relevant extra temporal jitter~\cite{strenzke_complexin-i_2009, yang_stochastic_2013}. However, this information can only be transmitted, if the outgoing APs were initiated with a sub-millisecond precision as well.
The timing precision is also relevant for phenomena such as spike timing-dependent plasticity~\cite{Dan2004}, and the spike time coding in sensory~\cite{Berry1997,VanRullen2005} and motor systems~\cite{Tang2014,Srivastava2017}.

It is known that a the population firing rate of pyramidal neurons can track changes in their input almost instantaneously. A population of layer 2/3 pyramidal neurons fires uncorrelated, when driven by fluctuating current, consistent with uncorrelated synaptic input. This is reminiscent of neuronal populations in-vivo~\cite{Ecker2010}. When a current step of 20 pA amplitude, as small as a single synaptic input, is added to each neuron's input simultaneously, the population firing rate significantly increases in less than a millisecond~\cite{tchumatchenko_ultrafast_2011,Ilin2013}. High temporal precision of AP initiation thus translates into ultrafast population encoding which is thought to be important for realizing fast information processing and decision making in the brain~\cite{Stanford2010}. %For example a highly demanding image discrimination task can be accomplished by humans within 150ms~\cite{Thorpe1996}, or, for macaque monkeys, it takes 150-200ms to execute a target saccade~\cite{Pare1996}. These require both the sensory and the motor pathways.
The underlying biophysical mechanisms of ultrafast population encoding and temporally precise firing of action potentials, however, are not well understood.

% relation between kink/rapid or sharp onset and bandwidth
Thus a key question in regards of the importance of the action potential waveform, specifically the onset rapidness, is whether and how it is linked to the encoding bandwidth.\\
The voltage dependence of the membrane current around the AP threshold, determines the AP onset rapidness at the initiation site, and the bandwidth of the dynamic gain: if the voltage dependence steeply increases around the threshold, onset rapidness and encoding bandwidth are high. However, it is not clear whether the observation of an AP high onset rapidness is sufficient to predict a high bandwidth. Such a predictive power would be very helpful. The AP waveform is comparably easy to measure but in itself not relevant for network function. Measuring the encoding bandwidth, on the other hand, requires thousands of spikes. However, the relation between AP onset rapidness and encoding bandwidth is  further complicated, if the AP waveform is observed at the soma, tens of micrometers away from the axonal initiation site.

The somatic AP waveform of cortical neurons has a high onset rapidness of around $30 \mathrm{ms}^{-1}$. The phase space plot of the somatic voltage rises very abruptly at the AP onset, producing a 'kink'. While this has been interpreted to predict an encoding bandwidth of several hundred Hertz~\cite{naundorf_unique_2006}, others have suggested the high somatic onset rapidness to result from a low onset rapidness at the AIS which is distorted by the propagation towards the soma~\cite{McCormick2007}. However, pyramidal neurons in acute brain slices and in-vivo have indeed been found to encode with a bandwidth of $200-400 \mathrm{Hz}$ ~\cite{higgs_conditional_2009,kondgen_dynamical_2008,naundorf_unique_2006,tchumatchenko_ultrafast_2011} or even $1 \mathrm{kHz}$ in the case of human cells~\cite{testa-silva_high_2014}. Furthermore, experimental studies found that a reduced axonal sodium current, either by partial sodium removal or by blocking of sodium channels, reduced the encoding bandwidth and the onset rapidness of the somatic AP~\cite{Ilin2013}.\\

Theoretical models can be coarsely categorized according to their morphology and AP generation. The simplest morphology is a point neuron which completely ignores all spatial structure~\cite{Brunel2001,fourcaud-trocme_how_2003,naundorf_unique_2006}, spatially extended neurons instead can account for the separation of AP initiation site from the soma, or effects of dendritic trees~\cite{brette_sharpness_2013,Telenczuk2017,Eyal2014,Ostojic2015}. AP generation in turn can be as simple as a threshold unit that produces an instantaneous spike as the potential threshold is crossed or be comprised of an AP dynamics. Surprisingly, the most simple model, a point neuron model with hard threshold AP generation, the leaky integrate-and-fire (LIF) model is capable of reproducing the encoding features of cortical neurons~\cite{Brunel2001}. Point neurons~\cite{fourcaud-trocme_how_2003,naundorf_unique_2006} and simple multi-compartment neurons~\cite{Huang2012,Wei2011} with biophysically motivated membrane currents or similar voltage dependence neither have a bandwidth far above the firing rate, nor do they display a pronounced change of dynamic gain when the input correlation time is changed. Thus the challenge to understand the observed phenomena is to isolate features that account for the observed dynamic gain in biophysically more realistic models.

% proposals of underlying biophysical cause for rapidness
Several hypotheses have been proposed to explain the high encoding bandwidth and the sudden onset of the somatic AP: i) cooperative gating of ion channels~\cite{naundorf_unique_2006}, ii) dendritic load on the soma~\cite{Eyal2014,Ostojic2015}, and iii) electrical decoupling between soma and AIS~\cite{brette_sharpness_2013,Telenczuk2017}. While i) argues for a mechanism directly affecting the AP dynamics, ii) and iii) propose a neuron morphology based argument. Specifically iii) propose an elegant approach based on a reduced multi-compartment model that provides novel insight in the possible origin of the somatic waveform 'kink'. Unaccounted for by single compartment neuron models, the AP initiation site of a cortical neuron is located in the AIS and, hence, separated from the soma~\cite{Stuart1994, Stuart1997}. This separation is proposed to cause the somatic membrane potential 'kink'. While it was suggested that this may also account for the observed input-output properties, a direct examination of the encoding ability has not been performed. 

% how to quantify fast population coding
To quantify the population encoding ability, the linear response function of the population firing rate has been adopted for experimental cell physiology. The linear response function represents the dynamic gain as a function of frequency. This concept was first proposed by Knight~\cite{Knight1972} and later theoretically elaborated in more biologically realistic and detailed neuron models~\cite{Knight1972a, Brunel2001, Lindner2001,fourcaud-trocme_how_2003, Fourcaud-Trocme2005, Naundorf2005,Tchumatchenko2011a, Wei2011, Huang2012, Touzel2015}.

% what we do in this study
In this work, we first reproduce the somatic 'kink' due to a more distal positioning of the AIS in a multi-compartment model similar to the one proposed in~\cite{brette_sharpness_2013}. We then determine the dynamic gain in the fluctuation driven regime injecting colored noise currents and investigate whether shifting the AIS along the axon does impact i) the bandwidth of the dynamic gain and ii) the sensitivity of the dynamic gain to the input correlation time. We furthermore explore the effects of different voltage sensitivities of the sodium activation as well as the impact of soma size on the linear input-output relation.

\section*{Materials and Methods}
\subsection*{Model}

All simulations were performed with NEURON 7.3~\cite{carnevale_neuron_2006} as a module for Python 2.6. The morphology, that is a ball-and-stick model with soma and axon, and its parameters are the same as in~\cite{brette_sharpness_2013} apart from a slight deviation from the ball-and-stick morphology by modeling the soma as a cylindrical compartment instead of a sphere (see Fig.~\ref{fig:sanitycheck}\textbf{A)}). The soma has a diameter of $d_S=50\mu m$ and length $l_S=50\mu{}\mathrm{m}$ while the axon is $d_A=1\mu{}\mathrm{m}$ thick and $l_A=600\mu{}\mathrm{m}$ long. The intracellular resistance is $R_a=150\Omega\mathrm{cm}$. The specific membrane capacitance is $c_m=0.75\mu{}\mathrm{F/cm}^2$ and the specific membrane resistance (inverse leak conductance) is $R_m=30000\Omega\mathrm{cm}^2$ with a leak reversal potential of $E_\mathrm{L}=-75 \mathrm{mV}$. For the axon, the passive currents are the same as for the soma. Sodium channels are modeled as a NEURON point process at the axon initial segment at distance $\x$ from the soma. As in~\cite{brette_sharpness_2013}, we only take the activation dynamics of the sodium conductance into account and omit inactivation. Peak sodium conductance $g_\mathrm{Na} = 5.23\cdot 10^{−3}\mathrm{S/cm}^2$. Since this simplified conductance model gives a dynamical system that does not produce a spike we define a spike threshold at $V=0$ and reset the voltage over the whole cell to $E_\mathrm{L}$ (which is equivalent to the resting potential for about the whole cell apart from the immediate neighborhood of the sodium current position) when the threshold is crossed, analogous to the spiking mechanism in integrate-and-fire neuron models~\cite{dayan_theoretical_2001}.

The neuron is characterized by the following partial differential equations. The voltage in the soma is
\begin{equation}
	c_m\frac{\partial V_S(x,t)}{\partial t} = \frac{l_S}{4R_a}\frac{\partial^2 V_S(x,t)}{\partial x^2} -\frac{V_S(x,t)-E_\mathrm{L}}{R_m} + \frac{I(t)}{\pi d_S}\delta\left(x-\frac{d_S}{2}\right)
    \label{eq:soma}
\end{equation}
with $I$ being the input current. The axonal membrane voltage is described by
\begin{equation}
	c_m\frac{\partial V_A(x,t)}{\partial t} = \frac{l_A}{4R_a}\frac{\partial^2 V_A(x,t)}{\partial x^2} -\frac{V_A(x,t)-E_\mathrm{L}}{R_m} - \frac{\bar{g}_\mathrm{Na}m(V_A(x,t))(V_A(x,t)-V_{\mathrm{Na}})}{\pi d_A}\delta\left(x-\x\right)
    \label{eq:axon}
\end{equation}
and
\begin{equation*}
    \tau_m \dot{m} = m_\infty(V) - m.
\end{equation*}
The boundary conditions of the equations~\eqref{eq:soma} and~\eqref{eq:axon} are given as $V_S(l_S,t)=V_A(0,t)$, $d_S\frac{\partial V_S(x,t)}{\partial x}|_{x=d_S}= d_A\frac{\partial V_A(x,t)}{\partial x}|_{x=0}$, $\frac{\partial V_S(x,t)}{\partial x}|_{x=0}=0$, and $\frac{\partial V_A(x,t)}{\partial x}|_{x=l_A}=0$.

\subsection*{Simulations}
% current
%% OU as input, how does lin. approx work?
We drove the model by injecting a Gaussian colored noise current $I(t)$ into the soma as done in experimental studies~\cite{kondgen_dynamical_2008,higgs_conditional_2009,boucsein_dynamical_2009,tchumatchenko_ultrafast_2011} and detailed by equation~\eqref{eq:soma}. We realize the colored noise by an Ornstein-Uhlenbeck process~\cite{tuckwell_stochastic_1989, gillespie_exact_1996} with correlation time $\tau$, standard deviation $\sigma$ and mean current of $\mu$:
\begin{equation*}
  \tau\mathrm{d}I(t) = (\mu - I(t))\mathrm{d}t + \sqrt{\tau}\sigma\mathrm{d}W(t)
\end{equation*}
where $W(t)$ denotes a Wiener process with zero mean and variance of one.
The numerical solution to the above model equations were computed by implicit Euler integration with temporal steps of $\mathrm{\Delta t}=25 \mu{}\mathrm{s}$ and spatial step sizes of $\mathrm{\Delta x}=1 \mu{}\mathrm{m}$.

Several conditions were maintained for all calculations to make results comparable. i) we kept the firing rate constant~\cite{fourcaud-trocme_how_2003} at $5\mathrm{Hz}$ similar to the experiments by Tchumatchenko et al.~\cite{tchumatchenko_ultrafast_2011}; ii) the neuron was driven in its linear response regime, that is the input current must not be too large which we achieve by demanding the standard deviation of the somatic membrane voltage to be about $5 \mathrm{mV}$; iii) we assured that the neuron is driven by current fluctuations rather than the mean current which is realized by guaranteeing that no spiking occurs with a constant, non-fluctuating input of the chosen mean current; and iv) we controlled for the coefficient of variation (CV) to be in the range $0.85\pm 0.05$ to make sure that the firing activity is not bursting and close to regular pyramidal cell activity.

\paragraph{Parameter space exploration.}
We explore the dependence of the dynamic gain on several parameters. Those are the input current correlation time $\tau \in \{5, 50\} \mathrm{ms}$, the position of the sodium currents $\x=\{20,40,80\}\mu\mathrm{m}$, the voltage sensitivity $k_a=\{0.1, 6\}\mathrm{mV}$~\footnote{The voltage sensitivity is a parameter of the activation function $m_\infty(V)=1/(1+\exp((V_{1/2}-V)/k_a)$.}, and the soma diameter $d_S=\{10, 50\}\mu\mathrm{m}$.

\subsection*{Computing the linear response of a neural ensemble}
% analysis
%% STA relation to power spectrum
In the linear response regime we can approximate the instantaneous firing rate of the ensemble of independent neurons by a linear filter acting on the input current. To analyze the bandwidth we thus need to compute the transfer function of this filter. While there are different ways to estimate the transfer function from spiking data and the stimulus current~\cite{kondgen_dynamical_2008,tchumatchenko_ultrafast_2011}, here we use the approach proposed in~\cite{higgs_conditional_2009} as outlined in what follows.% [WE FILTER IN THE COMPLEX SPACE RATHER THAN JUST THE REAL VALUED AMPLITUDE AS HIGGS AND SPAIN DO, I REMEMBER YOU TOLD ME WHY BUT I CAN'T RECALL THE ARGUMENT - CAN YOU? IF NOT I AM HAPPY TO LOOK INTO IT]  %Other approaches are vector or input currents that are which requires to compute the power spectrum of the input and the instantaneous firing rate.
The transfer function is given as
\begin{equation*}\label{eqn:transfer}
  G\left(f\right) = \frac{|\mathcal{F}\left(C_{I\nu}(\tau)\right)|}{|\mathcal{F}\left(C_{II}(\tau)\right)|}
\end{equation*}
where $\mathcal{F}$ denotes the Fourier transform, $C_{I\nu}$ and $C_{II}$ is the input-output correlation and input auto-correlation function, respectively. The Fourier transform of an auto-correlation function is the power spectrum according to the Wiener-Khinchin theorem whose analytical form is known for an Ornstein-Uhlenbeck process as $\mathcal{F}(C_{II}(\tau))=P(f)=\frac{2\tau\sigma^2}{1+((2\pi f)^2}$.\\
To compute input output correlation function we make use of the following equality
\begin{equation*}
  C_{I\nu}(\tau) = \langle \delta I(t-\tau) \delta \nu(t) \rangle = \frac{1}{NT}\int
  \mathrm{d}t\delta I(t-\tau)\sum_{i,j}\delta(t-t_i^j)
\end{equation*}
$\delta(\cdot)$ is the Dirac delta, $\delta \nu$ and $\delta I$ denote small deviations. This illustrates that we can conveniently get $C_{I\nu}(\tau)$ by computing the spike-triggered average (STA).
For computing the STA we need to select a time window large enough for correlations to be decay, here we choose a window of $800 \mathrm{ms}$ centered about the spike time. We then computed the Fourier transform and applied a filter bank of Gaussians in the complex plane to de-noise before computing the power of the signal by taking the absolute value.

We then performed this computation of the dynamic gain for each parameter value by running a numerical simulation of the stochastic current driven model described above for 20.5s (with 0.5s of initialization to make sure the initial conditions of the model don't affect the results and 20s of actual recording). Given the 5Hz firing rate we gather on average 100 spikes per run. We repeat this simulation 50 times and compute a single STA from the resulting time series.\\
In order to estimate the regime of statistical accuracy of our simulation we perform two analyses. First we compute a confidence interval by bootstrapping and second we compare the result to the null hypothesis that the transfer function stems from random spike times.
Bootstrapping is achieved simply by choosing 400 samples with replacement from the set of 400 spike triggered current time series to compute the STA. We do this 1000 times and from those boot strapped transfer functions we take the upper and lower bound define the 95\% confidence interval around the true transfer function.
For the null hypothesis testing we need to shuffle the spike times with the constraint of keeping the firing rate and ISI distribution. In order to do so we add a random number to all spike times produced by a single simulation run of 20 seconds. A different number is added to each of the 50 repetitions and the spike times are computed modulo 20 to assure they don't exceed the absolute time window. We then repeat this procedure 500 times and take the upper bound of the resulting 95\% interval as the curve corresponding to the null hypothesis. We took this curve as delineation of validity of our numerically computed dynamic gain functions and show only the valid regime above this significance border.

% Results and Discussion can be combined.
\section*{Results}
\begin{figure}
\includegraphics[width=\textwidth]{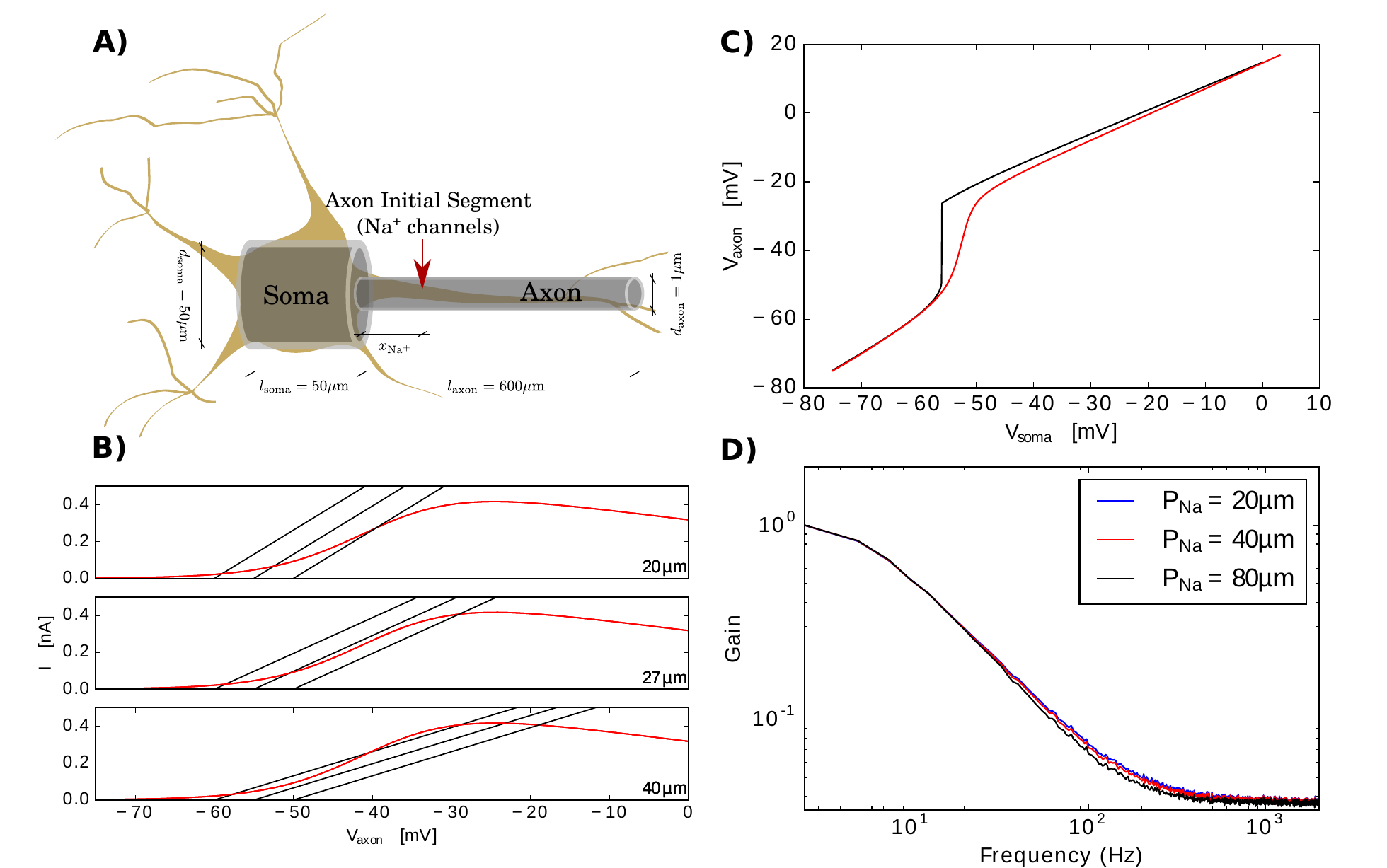}
\caption{\textbf{Model properties}. \textbf{A)} A simple sketch of the model morphology: both soma and axon are modeled as cylinders. The sodium current is positioned at $x_\mathrm{Na}$ and the stimulus current is injected into the center of the soma. More simulation and model details are given in the main text. \textbf{B)} We reproduce the bifurcation property discussed in Fig. 2 of~\cite{brette_sharpness_2013}. Sodium current (red) and lateral current (black) is plotted as function of the voltage at the axon initial segment for different positions of the AIS (top to bottom: $x=20\mathrm{\mu m}, 27\mathrm{\mu m}, 40\mathrm{\mu m}$). \textbf{C)} Voltage at the axon initial segment as a function of somatic voltage. Patch clamped (black) curve in comparison with the dynamic case (red). Analogous to Fig. 1F of~\cite{brette_sharpness_2013}, the patch clamped case shows a loss of voltage control around $-55\;\mathrm{mV}$. However, we point out that the dynamic case shows no sharp, localized transition. This hints that the loss of voltage control is less pronounced in the dynamic case, that is during an action potential. \textbf{D)} Gain is plotted as a function of input current frequency for the sub-threshold responses to a colored noise stimulus with correlation time $\tau=5\mathrm{ms}$. We investigate the sub-threshold transfer function for three different positions of sodium channels, colors denoted in the legend.
}
\label{fig:sanitycheck}
\end{figure}

\begin{figure}
\includegraphics[width=\textwidth]{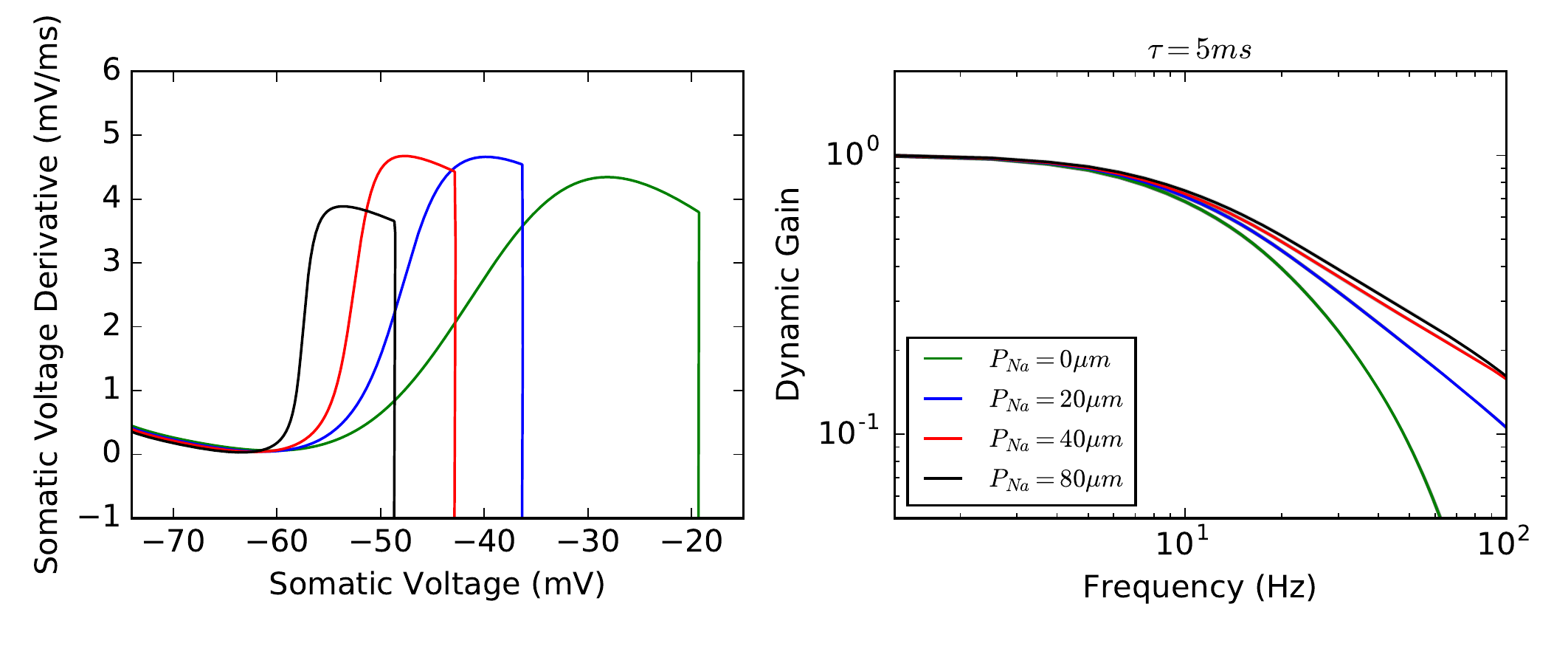}
\caption{\textbf{Phase plot and transfer functions.} \textbf{A)} Phase plot for the soma, i.e. we plot time derivative of the voltage over the voltage. Colors correspond to different sodium channel positions $x_\mathrm{Na}$ (confront legend in B) ). The vertical straight line results from the artificial resetting mechanism in this model. \textbf{B)} Comparison of transfer functions for different positions of the sodium current. The input current correlation time is set to $\tau=5\mathrm{ms}$. All curves are normalized such as to start with a gain of one for better comparison of the dynamic gain's tail. 95\% confidence intervals are plotted as shaded areas, however, for the large part of the transfer function these are overlapping with the average curve. All curves are above statistical significance (detailed explanation in {\em Material and Methods}).  %[NOTE: extend the frequency axis to 500Hz? Then the next sentence makes sense] Shaded areas surrounding the transfer functions are the 95\% bootstrapped boundaries. %Transfer functions for a multi-compartment model with sodium current at the intersection of the axon and soma, i.e. $x=0\mu\mathrm{m}$, compared to a single compartment model (purple) that shares the passive properties with the multi-compartment model's soma, with an added sodium current in addition. B) 
%C: some plot that clarifies that continuous spiking is happening if distance $>$ lambda. \\ We plot transfer functions for different realizations of the parameters $\tau$ and $x$. (A) and (B) show how transfer functions are affected when we change the sodium channel position while keeping all other parameters fixed. Specifically, in (A) we plot the results for $\tau=5 \mathrm{ms}$ and (B) for $\tau=20 \mathrm{ms}$, the legend gives the different sodium channel positions corresponding to the line colors. 
}
\label{fig:transfer_PosNa}
\end{figure}

We built a two compartment model (Fig~\ref{fig:sanitycheck} \textbf{A)}) with identical characteristics as proposed in Brette 2013~\cite{brette_sharpness_2013} (see {\em Material and Methods}). To make sure that our model compares to Brette's in its central features we show the bifurcation of the voltage at the soma in dependence of the sodium channel's position. As can be seen in Fig~\ref{fig:sanitycheck} \textbf{B)} we obtain the same result as shown in Fig. 2 in Brette (2013)~\cite{brette_sharpness_2013}. We further reproduced the decoupling observed in Fig. 1F of~\cite{brette_sharpness_2013} as demonstrated in Fig~\ref{fig:sanitycheck} \textbf{C)}. While we found the same loss of voltage control about -55 mV between the two compartments when clamping and slowly increasing the somatic voltage, we point out that in the dynamic case (i.e. no clamp is applied) the decoupling of the compartments is less pronounced. We further investigated the sharpness of the shape of the somatic phase plot in dependence of the sodium current position~\ref{fig:transfer_PosNa}\textbf{A)}. We found that the speed by which the voltage rises increases the further distant the sodium channels are from the soma. However, the key quantity to explore fast population encoding, the dynamic gain, had not been investigated for this multi compartment model yet. We thus performed a thorough analysis of the dynamic gain in dependence of several different model parameters in the following. \\

% In line with prior work

\paragraph*{Somatic kink is not sufficient for a high information bandwidth.}
First we asked whether the dynamic gain depends on the position of the sodium channels. As can be seen in Fig.~\ref{fig:transfer_PosNa} \textbf{B)} the gain curves do not show a higher cut-off frequency when the position of sodium channels is moved further away from the soma. The low bandwidth we observe stands in contrast with previous experimental findings demonstrating that pyramidal neurons do have a cut-off frequency in the regime of hundreds of Hz. Instead the cut-off frequency is the same as previously found in conductance based single compartment models~\cite{fourcaud-trocme_how_2003}.

\begin{figure}
\includegraphics[width=\textwidth]{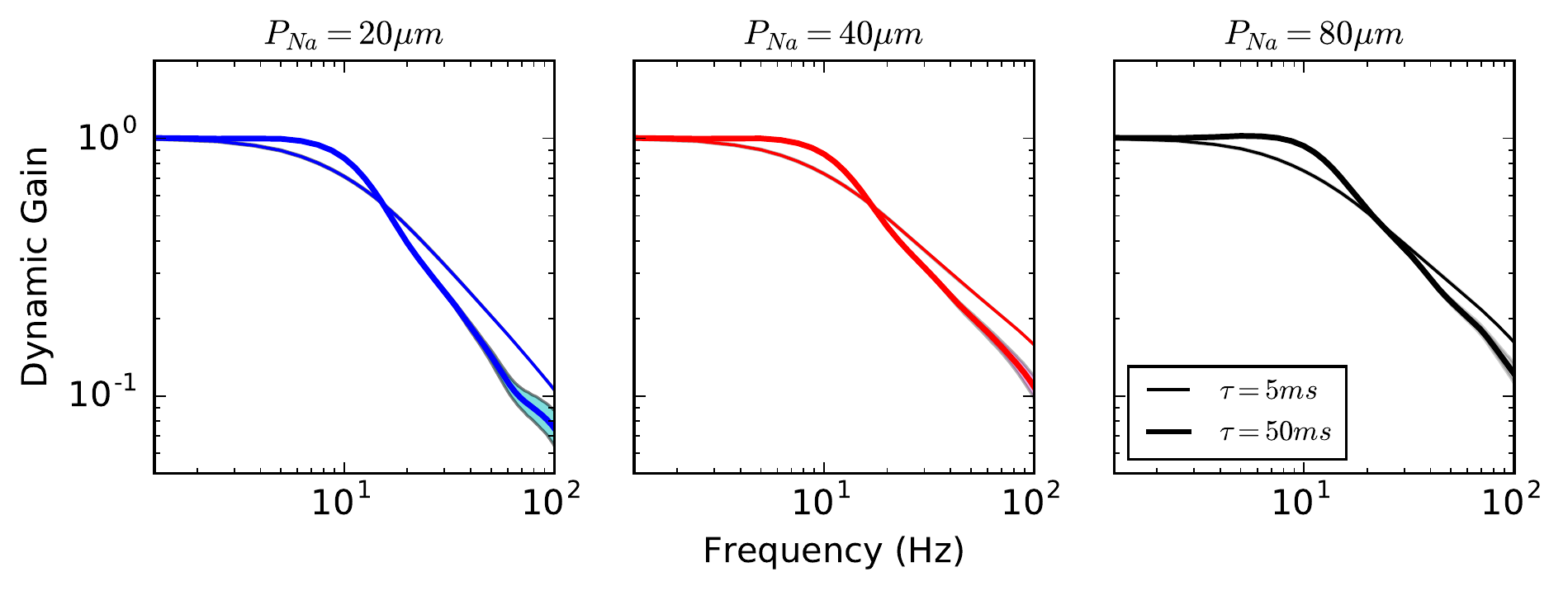}
\caption{\textbf{Effect of input correlation time on dynamic gain.} We plot the transfer functions for different input current correlation times to investigate whether the model shows dependence of the input correlation time, also called the Brunel effect~\cite{Brunel2001}. Model properties are the same as for Fig.~\ref{fig:transfer_PosNa}. Graph colors code different sodium channel positions with the same color code as in Fig.~\ref{fig:transfer_PosNa}, the distance increases from the left to the right panel as $x_{\mathrm{Na}}=20, 40$ and $80\mu\mathrm{m}$. Thin lines correspond to small correlation times ($\tau=5\mathrm{ms}$) and thick lines correspond to large correlation times ($\tau=50\mathrm{ms}$)}
\label{fig:BrunelEffect}
\end{figure}

\paragraph*{Input correlations don't alter bandwidth.} Stimulating the neuron with colored noise current simulates the barrage of synaptic input to a neuron with the current correlation time reflecting an effective synaptic time constant. Previous studies have shown that in simple model neurons as the LIF and other models with a large onset rapidness show sensitivity to this input correlation time. More recently experimental evidence has been provided in support of this effect~\cite{tchumatchenko_ultrafast_2011}. We find that the present ball-and-stick multi-compartment conductance based model fails to reproduce this dependence on input correlations for any of the different sodium channel positions (Fig.~\ref{fig:BrunelEffect}).

\begin{figure}
\includegraphics[width=\textwidth]{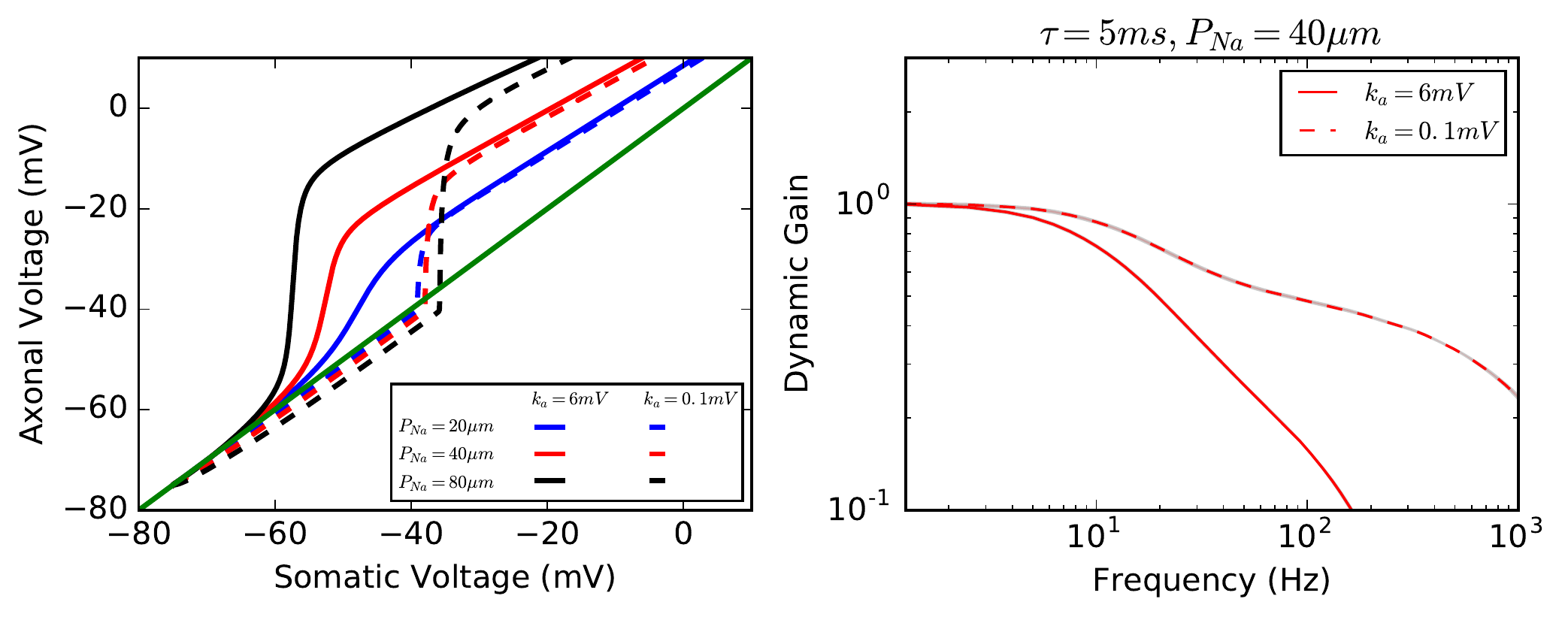}
\caption{\textbf{High voltage sensitivity.} \textbf{A)} For different sodium current positions (confront legend), we plot the axonal voltage at the position of the sodium currents over the somatic voltage. Solid lines are voltage traces for low voltage sensitivity $k_a=6$, dashed lines denote the same traces for high voltage sensitivity $k_a=0.1$. \textbf{B)} We compare transfer functions for different voltage sensitivities. }
\label{fig:volt_sens1}
\end{figure}

\begin{figure}
\includegraphics[width=\textwidth]{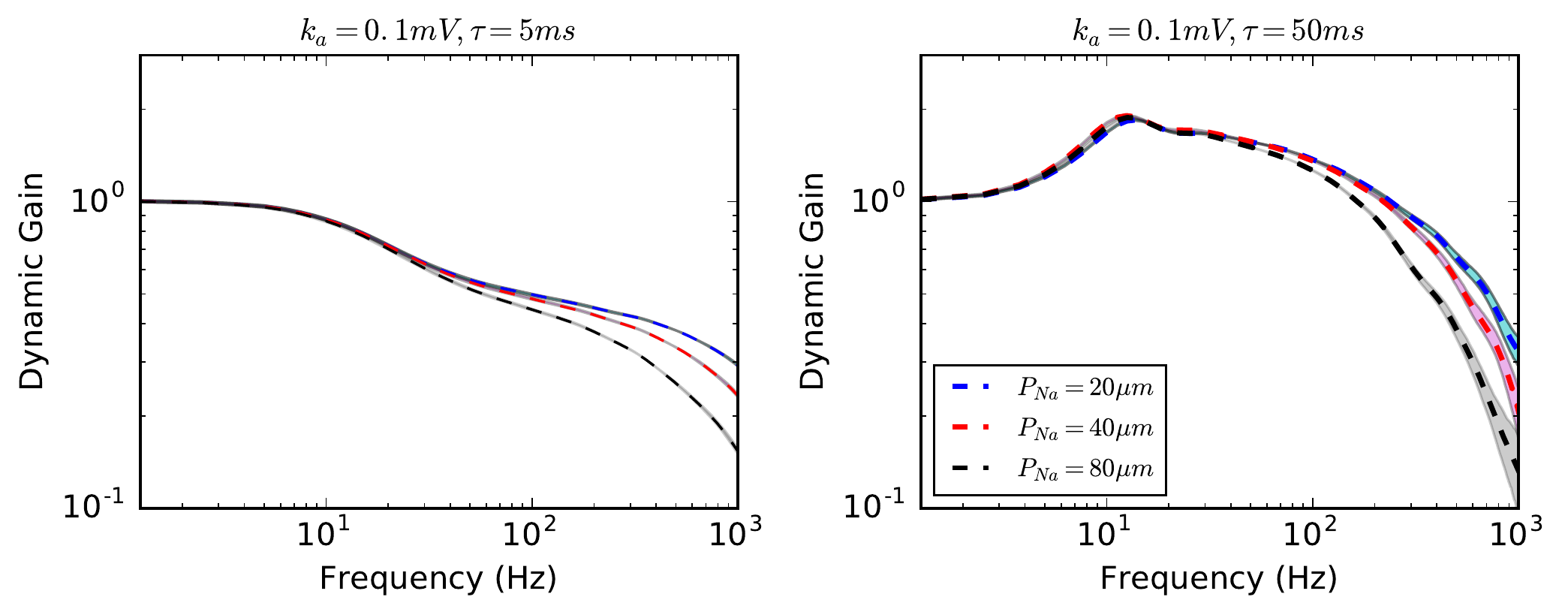}
\caption{\textbf{High voltage sensitivity and sodium channels position.} Dynamic gain curves are plotted for \textbf{A)} a short input current correlation time of $\tau=5\mathrm{ms}$ and \textbf{B)} a long correlation time of $\tau=50\mathrm{ms}$. We plot dashed lines to consistently denote linear response functions for models with high voltage sensitivity. Shaded areas define the 95\% confidence interval. Colors denote different sodium current positions as explained in the legend. }
\label{fig:volt_sens2}
\end{figure}

\paragraph*{High voltage sensitivity recovers empirical findings.}
We have shown that somatic onset rapidness does not imply fast population coding as quantified by the dynamic gain. These empirical findings, however, can be recovered when changing the voltage sensitivity parameter $k_a$ from $6\mathrm{mV}$ to $0.1\mathrm{mV}$, that is, making the sodium current activation curve steeper and, hence, the neuron model more voltage sensitive at the AIS. In Fig.~\ref{fig:volt_sens1}\textbf{A)} we show that i) the compartment decoupling becomes more pronounced for all sodium channel positions tested if voltage sensitivity is increased and ii) We note that these voltage plots are recorded in the dynamic regime, not by performing a static voltage clamp measurement. Importantly, Fig.~\ref{fig:volt_sens1}\textbf{B)} shows that the dynamic gain decays slower and, hence, the information bandwidth is increased as had been reported for single compartment models in prior work~\cite{fourcaud-trocme_how_2003,Naundorf2005,Wei2011}.\\
Next we explored the effect of sodium channel position on the dynamic gain for the model with high voltage sensitivity and found that there is no substantial difference in terms of information bandwidth between the conditions (Fig.~\ref{fig:volt_sens2}). Furthermore, we compared different input current correlation times and found that higher correlation times do increase the information bandwidth as had first been found by Brunel, et al. for the leaky integrate and fire neuron~\cite{Brunel2001}.

% quote brette "The mathematics of this phenomenon are very close to the cooperativity model, "  although directly implying that the compartmentalization also explains high bandwidth, we find that this is not the case.

%- this bifurcation is less clear in the dynamical picture when the soma also depolarizes during the AP. This makes it unclear whether there is a bifurcation at all. (For this we have to read Brette's second paper about this. Maybe he quantifies it better there.

\begin{figure}
\includegraphics[width=\textwidth]{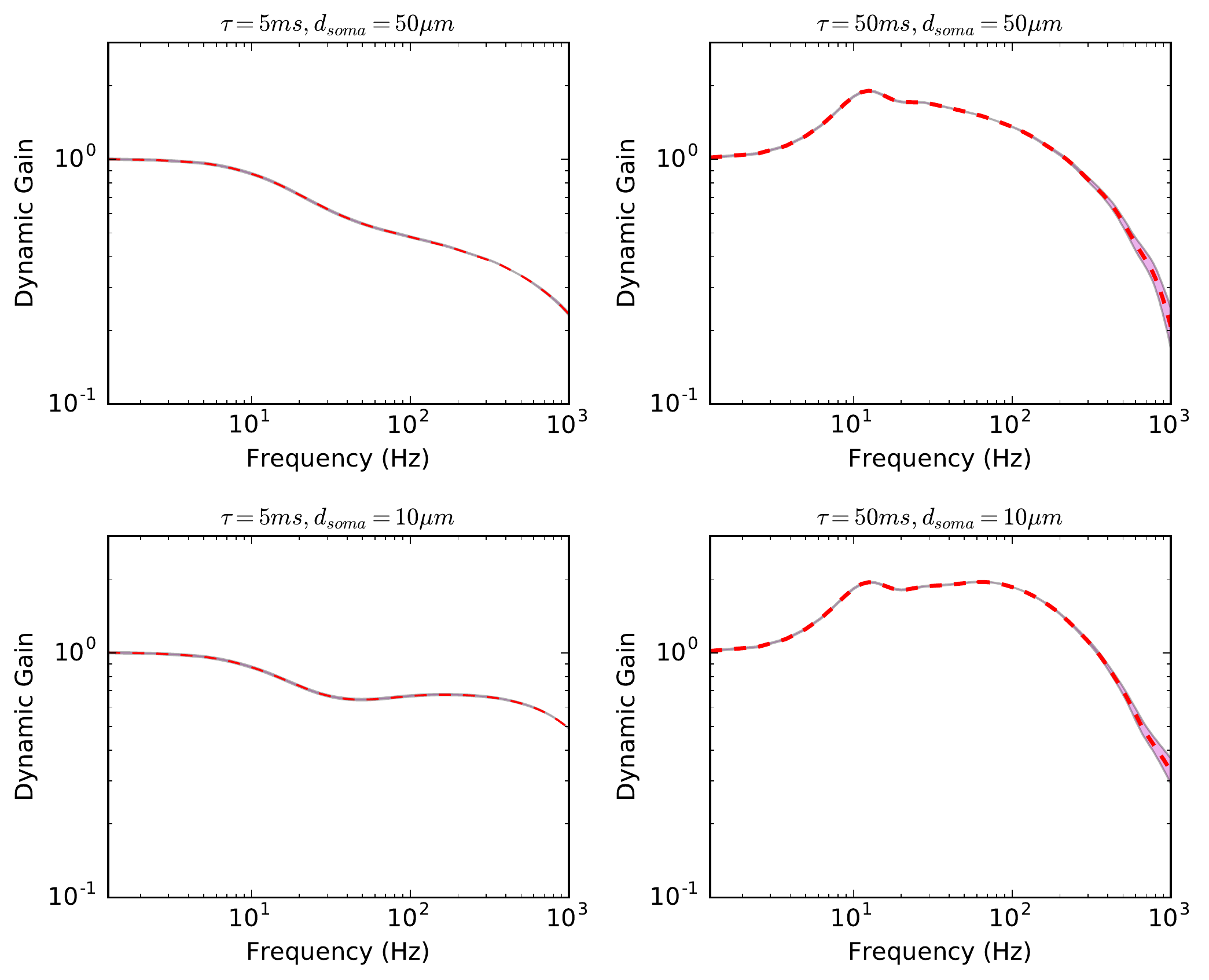}
\caption{\textbf{Effect of soma size.} We investigate the soma size effect on the linear transfer function for different scenarios (confront plot titles). As in previous plots, dashed lines denote high voltage sensitivity and shaded areas the 95\% confidence interval.}
\label{fig:somasize}
\end{figure}

\paragraph*{Soma size affects the dynamic gain for the highly voltage sensitive model.}
The last parameter that we tested was the somatic diameter $d_\mathrm{S}$. As $50\mu\mathrm{m}$ is a reasonable upper limit for cortical neurons we tested $10\mu\mathrm{m}$ which demarcates the lower end of the cortical soma size spectrum~\cite{dayan_theoretical_2001}. Since for low voltage sensitivity we could not find any significant changes when reducing the soma size, in Fig.~\ref{fig:somasize} we show the linear response functions for varying soma sizes in the case of high voltage sensitivity. We find that reducing the soma size increases the bandwidth effectively bringing the transfer functions to resemble more those of a LIF. Indeed this is not surprising given that we approach the limit where our two compartment model becomes a single compartment model whose response function characteristics are well studied. However, our results show that our findings are indeed robust with respect to soma size and don't depend on an artificially large soma as compared to most cortical cells.

\section*{Discussion}
Understanding the biophysics of neuronal action potentials has a long history with substantial successes in modeling and experiment~\cite{Hodgkin1952,Hamill1981} and is fundamental to our understanding of neuronal information processing. An aspect of current interest are the features of the neuronal action potential and the morphological and electrical organization of nerve cells that can enable high bandwidth ultrafast encoding. Interestingly, a substantial fraction of models fail to exhibit a realistically high cut-off frequency of hundreds of Hz of a population of neurons~\cite{higgs_conditional_2009,tchumatchenko_ultrafast_2011} as well as the experimentally observed sensitivity to the input correlation time of the dynamic gain~\cite{tchumatchenko_ultrafast_2011} while a simplistic neuron model like the LIF can account for those features~\cite{Brunel2001}. However, due their abstract nature and simplicity such models do not allow us to further investigate the biophysical features that are responsible for these measurable phenomena. Several hypotheses have been put forward as possible biophysical explanations: i) ion channel cooperativity~\cite{naundorf_unique_2006}, ii) dendritic load~\cite{Eyal2014,Ostojic2015}, and iii) compartmentalization of soma and axon through more distal position of the AIS~\cite{brette_sharpness_2013,Telenczuk2017}.\\

In this study we investigated the model proposed in~\cite{brette_sharpness_2013}, a simplified and analytically tractable conductance based multi-compartment model that was proposed to explain the rapid action potential onset~\cite{brette_sharpness_2013,Telenczuk2017}. The key feature of this model is that rapid onset in the somatic membrane potential can be achieved by moving the position of the AIS further away form the soma and thus compartmentalizing the neuron, viz. electrically decoupling the soma and axon.\\

Interestingly, a series of experimental studies have shown that the AIS can undergo plastic changes due to spiking activity and specifically relocate its position along the axon~\cite{Kuba2006,Kuba2010,Kuba2015,Grubb2010,Grubb2011}. These studies raise the question as to whether AIS position plays a role in neural coding. We investigated whether the position of the AIS can impact the linear response of a population of neurons. We employed the simplified ball-and-stick neuron model that was proposed in~\cite{brette_sharpness_2013} and investigated whether relocation of the AIS can reproduce the two key experimental findings: high cut-off frequency of the dynamic gain and input correlation time dependence of the dynamic gain. In doing so we also shed light on the relationship between rapid onset and large bandwidth.\\

While we reproduced the somatic rapid AP onset for distal positions of the AIS as shown in~\cite{brette_sharpness_2013} we found that the linear response remains unaffected by the specific position of the AIS. Furthermore, we found that the linear response of this model is insensitive to changes in input correlation time. Besides revealing that this reduced model cannot account for ultrafast neural responses as found in experiments, our findings also highlight that a rapid onset at the soma is not sufficient for producing a large encoding bandwidth. We show that one way to reconcile experimental results with this model is by tuning the activation slope of the sodium activation function which can be done by increasing the voltage sensitivity of the sodium channels. While we get more realistic cut-off frequencies the results, however, are still not sensitive to the position of the AIS.\\

We conclude thus that first our study suggest that rapid onset at the action potential initiation site can be linked to rapid responses of a neuron ensemble but not somatic rapid onsets since those can be caused by mechanisms not responsible for the high neural bandwidth. However, we point out that reliably measuring AIS membrane potentials pose an outstanding experimental challenge that has yet to be solved~\cite{Hu2009, Dulla2009}. Second, pure dislocation of the AIS does not affect population coding. This is a prediction that can be tested experimentally to further elucidate the biophysical factors underlying high bandwidth ultrafast population encoding. On the one hand, if the prediction is correct models need not precisely account for this morphological aspect when studying population dynamics. However, other morphological factors like the dendritic tree size have been shown to be relevant~\cite{Eyal2014,Ostojic2015} and thus further theoretical investigation is needed to identify an adequate minimal model that can guide the identification of the biophysical properties that determine dynamic population coding. %On the other hand, if the dynamic gain is indeed affected by the AIS position it would imply that this minimal model needs to be augmented with missing key features. However, we note that the linear response is by no means exhaustive of neural coding properties and thus more experimental and theoretical work will be necessary to elucidate the possible functions of AIS plasticity.

\section*{Acknowledgments}
This work was partially supported by the Federal Ministry for Education and Research (BMBF) under grant no. 01GQ1005B (to A.P., T.G., F.W. and D.B.), through CRC 889 by the Deutsche Forschungsgemeinschaft and by the VolkswagenStiftung under grant no. ZN2632 (to F.W.), and by the China Scholarship Council. We thank Rainer Engelken and
Barbara Feulner for fruitful discussions.

%\nolinenumbers

% Compile your BiBTeX database using our plos2015.bst
% style file and paste the contents of your .bbl file
% here.

\bibliography{bibliography}

\end{document}